\documentclass[a4paper,11pt]{article}
\pdfoutput=1 

\usepackage{jinstpub} 

\usepackage{lineno}

\usepackage{caption}
\usepackage{subcaption}

\title{\boldmath Status and recent extensions of the Caribou DAQ System for picosecond timing}

\author{E. Buschmann}
\affiliation{CERN,\\Meyrin, Switzerland}

\emailAdd{eric.buschmann@cern.ch}

\abstract{
  Caribou is a flexible open-source DAQ system developed and used within several collaborative frameworks (CERN EP R\&D, RD50, AIDAinnova) for laboratory and high-rate beam tests and easy integration of new silicon-pixel detector prototypes. It uses common hardware, firmware and software components that are shared across different projects, thereby reducing the development effort and cost for such readout systems significantly.

  This contribution introduces the FASTPIX, APTS and DPTS detectors recently integrated in Caribou and their requirements in terms of waveform sampling and timing of digital pulses.
  A Time-to-Digital-Converter implemented in the Caribou FPGA was developed to overcome the limitations of the previously used oscilloscope-based readout. It allows for precision timing over large time spans required for the Time-of-Arrival, Time-over-Threshold, and position encoding used in the FASTPIX and DPTS test chips with asynchronous digital readout. A time resolution after calibration of better than 10\,ps is achieved.
}

\keywords{Data acquisition concepts, Timing detectors}

\proceeding{Topical Workshop on Electronics for Particle Physics - TWEPP2022\\
  19-23 September 2022 \\
  Bergen, Norway}

\begin{document}
\maketitle
\flushbottom

\section{Introduction}

\begin{figure}[htbp]
  \centering
  \includegraphics[width=0.95\textwidth]{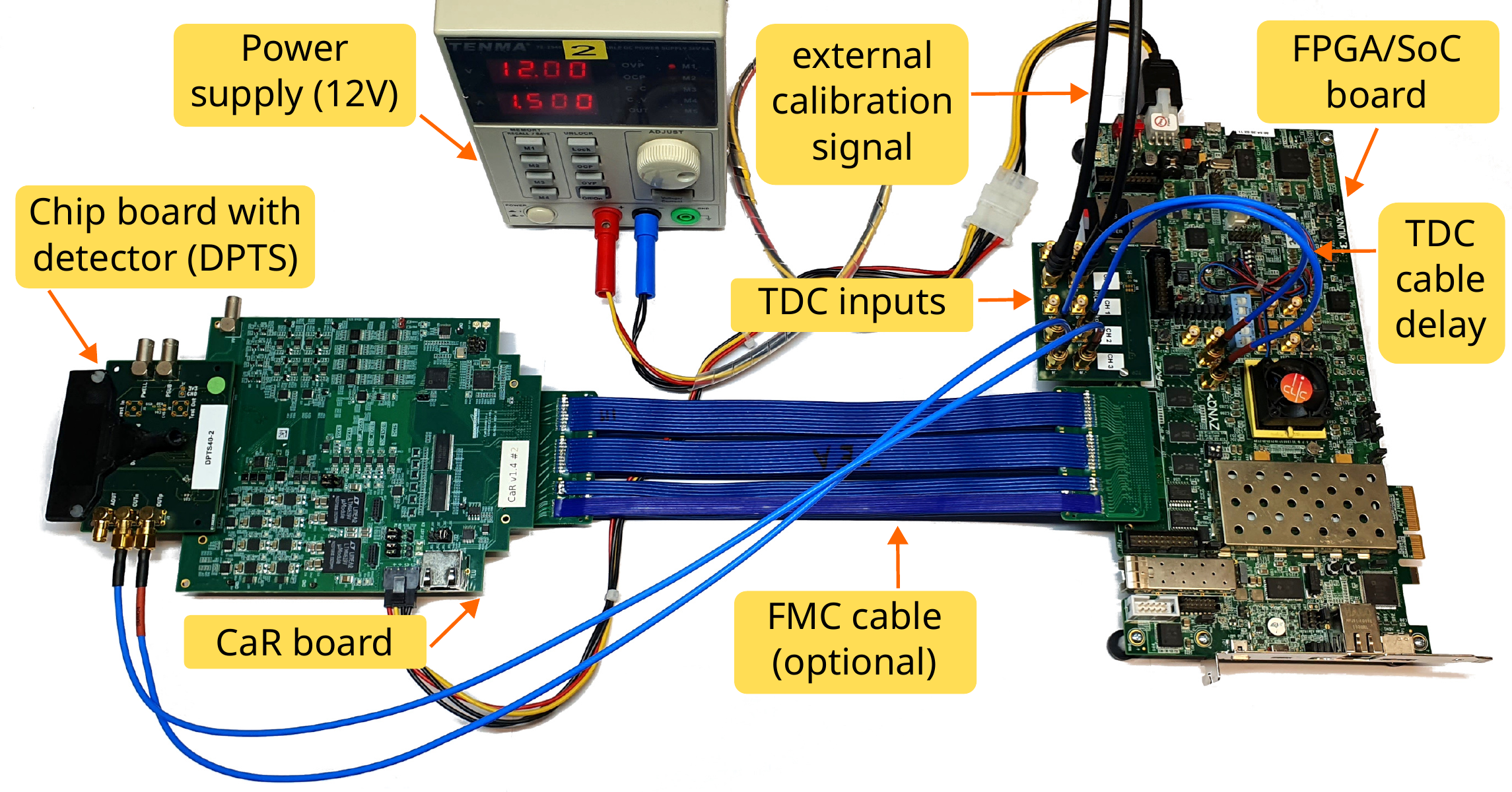}

  \caption{\label{fig:caribou} Caribou setup.}
\end{figure}

Caribou is a modular readout system that reduces the effort required for the integration of new detector prototypes \cite{Caribou, Caribou2}.
Figure~\ref{fig:caribou} shows the components of a typical Caribou setup and the recently added extensions presented here. It consists of a Xilinx ZC706 board which runs the Peary readout software and detector-specific firmware and is connected to the CaR board, which provides a number of resources such as power supplies, reference voltages and currents and interfaces to the detector-specific chip board, which houses the detector.

\section{New detectors and DAQ requirements}

The FASTPIX, DPTS and APTS chips were recently integrated into Caribou.
FASTPIX is a monolithic pixel sensor demonstrator with sub-nanosecond time resolution implemented in a modified 180 nm CMOS imaging process \cite{FASTPIX,FASTPIX2}.
FASTPIX has 32 matrices with 4x16 hexagonal pixels each. The matrices differ in pixel pitch (8.6\,µm, 10\,µm, 15\,µm, 20\,µm) and other layout and processing parameters.
DPTS and APTS are monolithic pixel sensors and technology demonstrators manufactured in a modified 65\,nm CMOS imaging process \cite{DPTS}. The DPTS has a $32\times32$ pixel matrix with $15\times15\,\textrm{µm}^2$ square pixels. The APTS has a $4\times4$ pixel matrix with a pitch of 10\,µm to 25\,µm.
Readout of the 16 analog channels of the APTS is achieved using a 16 channel ADC with 14-bit resolution at 65 MSPS on the CaR board, which was integrated in Peary for beam-test measurements.

\subsection{Timing requirements}

\begin{figure}[htbp]
  \centering
  \includegraphics[width=.65\textwidth]{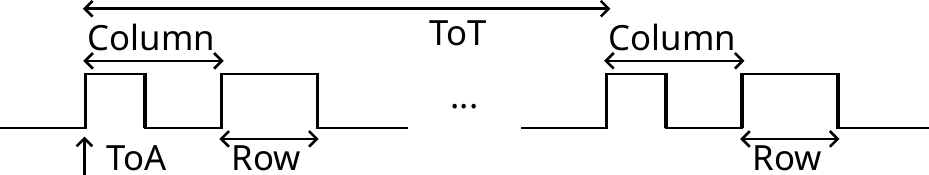}

  \caption{\label{fig:decoding} Sketch of the delay-based encoding scheme used by the DPTS. Shown is the sequence of pulses originating from a single pixel hit.}
\end{figure}

Both FASTPIX and DPTS use a similar asynchronous digital readout scheme where the pixel position, Time-of-Arrival (ToA), and Time-over-Threshold (ToT) are encoded in the relative timing of signal edges using on-chip delay lines. As this is not synchronised to a clock, special care must be taken when sampling the signal to properly record the timing information.
Figure~\ref{fig:decoding} illustrates the position and ToT encoding used by the DPTS. Each pixel hit is encoded in 2 pairs of pulses where the time difference between the first and the second rising edge encodes the column and the difference between the second rising and the second falling edge encodes the row. The ToT is the time difference between the rising edges of the first and second group. The column and row information is repeated in the second group to be able to match the correct pairs in events with more than one pixel hit.
The separation between pixels in column and row is $\mathcal{O}(100\,\textrm{ps})$ with $t_{column}$ and $t_{row}$ of $\mathcal{O}(1\,\textrm{ns})$ and a ToT of up to $\mathcal{O}(10\,\textrm{µs})$.
This is especially challenging for an oscilloscope-based readout, as it requires recording tens of microseconds of data at a sufficiently high sampling rate, which has been the limiting factor in previous test-beam campaigns in terms of achievable trigger rate.

\section{Time to Digital Converter}
\subsection{TDC requirements}

A time-to-digital converter (TDC) was developed for the Caribou system, geared towards the readout of the FASTPIX and DPTS demonstrators but also suitable for timing measurements in general.
The DPTS encoding scheme gives a number of requirements for the TDC.
The events consist of several rising and falling edges within a few nanoseconds, which means that the TDC should have no dead time as not to cut off any part of an event.
It should be able to record both rising and falling edges and especially several edges within one TDC clock cycle.
The TDC should also be able to resolve edges with less than 1 ns separation to be able to record short pulses or several pulses close together that can arise under some conditions.
To be able to distinguish the pixels, a time resolution of better than about 100\,ps is required.
If the TDC is not only used for decoding but also to measure the time resolution of a device under test, an even better time resolution is desirable.

\begin{figure}[htbp]
  \centering
  \includegraphics[width=.75\textwidth]{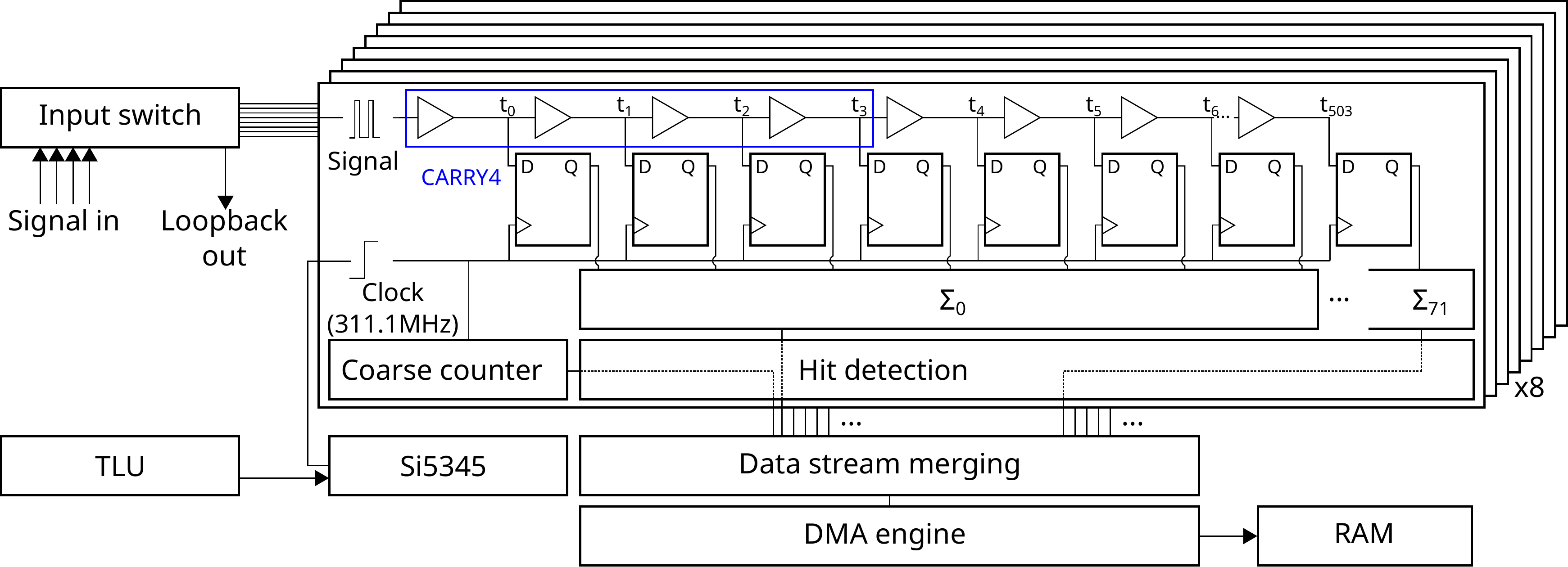}

  \caption{\label{fig:tdc} Block diagram of the 8 channel TDC.}
\end{figure}

\subsection{TDC concept}

An 8 channel TDC prototype which uses a direct readout of the delay line was developed to demonstrate the feasibility for this application. Figure~\ref{fig:tdc} shows a schematic view of the TDC architecture, which is implemented in the Zynq programmable logic of the Caribou system.
The TDC uses tapped delay lines composed of CARRY4 logic elements \cite{xilinx}, which have dedicated routing between elements and allow for better control of routing delays. Each delay line consists of 126 CARRY4 elements with 4 taps each for a total of 504 taps. The expected length of the delay line is roughly between 4\,ns and 7\,ns, which is extracted from the design for the slowest and fastest process corner.
The sampling registers are constrained to be placed in the same logic slice as the CARRY4 element. No other constraints are applied.
Taps are merged and summed in groups of 7 to reduce the number of output bits from 504 to 216. Each TDC conversion produces a 256\,bit wide vector, which leaves 40\,bit for the coarse counter, channel number and other status flags.
A hit detection logic is employed to detect signal edges in the data and forward it to the readout. This is also triggered on overflows of the coarse counter to extend the range of the counter.
The data streams from all TDC channels are merged and written via DMA to memory from where they are retrieved by the readout software and stored for later analysis.

The Si5345 clock generator on the CaR board is used as a low-jitter 311.1111\,MHz clock source for the TDC, which allows to synchronise the clock with a trigger logic unit for consistent timestamps across multiple devices such as a beam telescope setup. The highest clock frequency that still achieves timing closure was chosen.

\subsection{Calibration \& Characterisation}

\begin{figure}[htbp]
  \centering

  \begin{subfigure}[b]{0.49\textwidth}
    \centering
    \includegraphics[width=\textwidth]{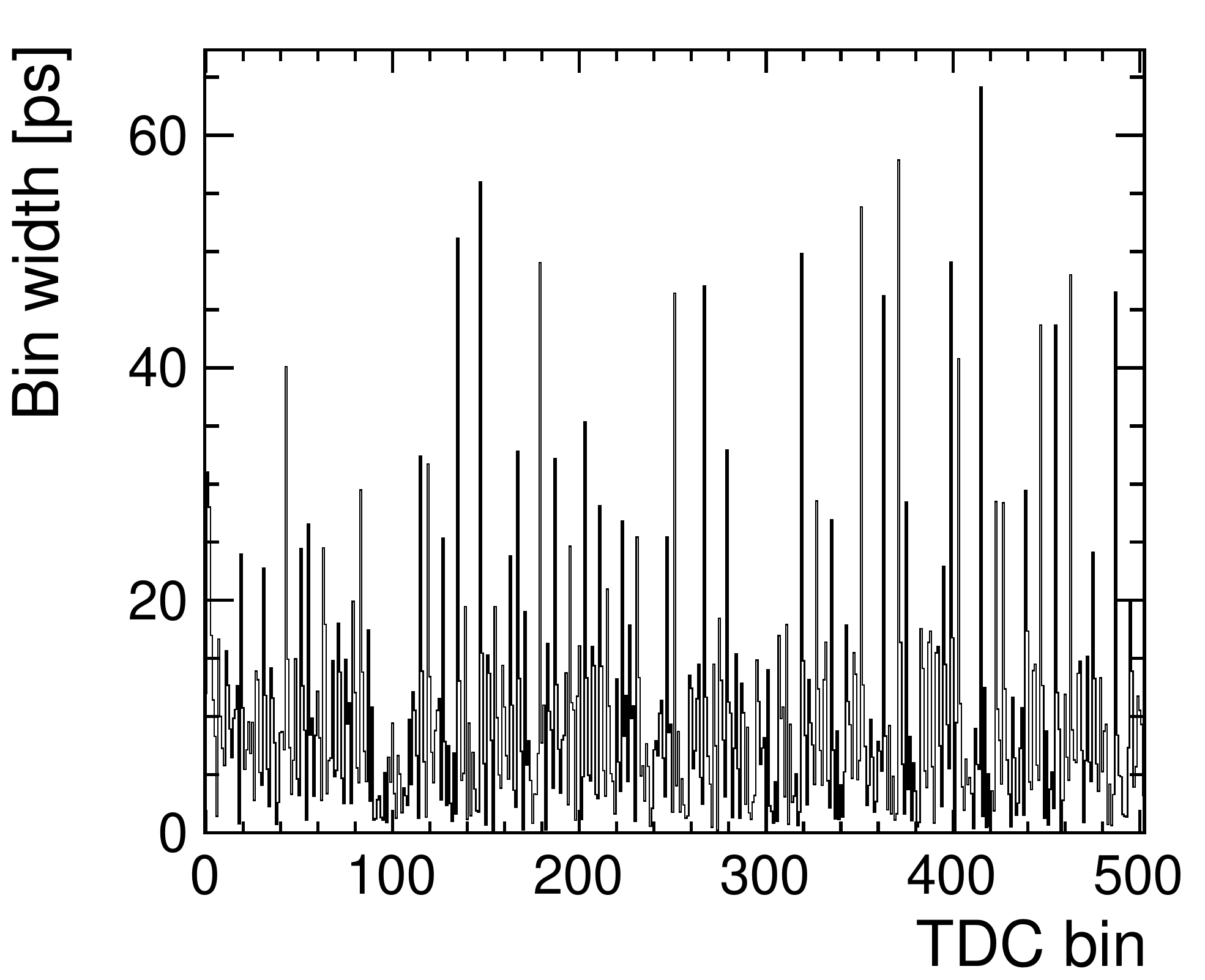}
    \caption{}
    \label{fig:bins}
  \end{subfigure}
  \begin{subfigure}[b]{0.49\textwidth}
    \centering
    \includegraphics[width=\textwidth]{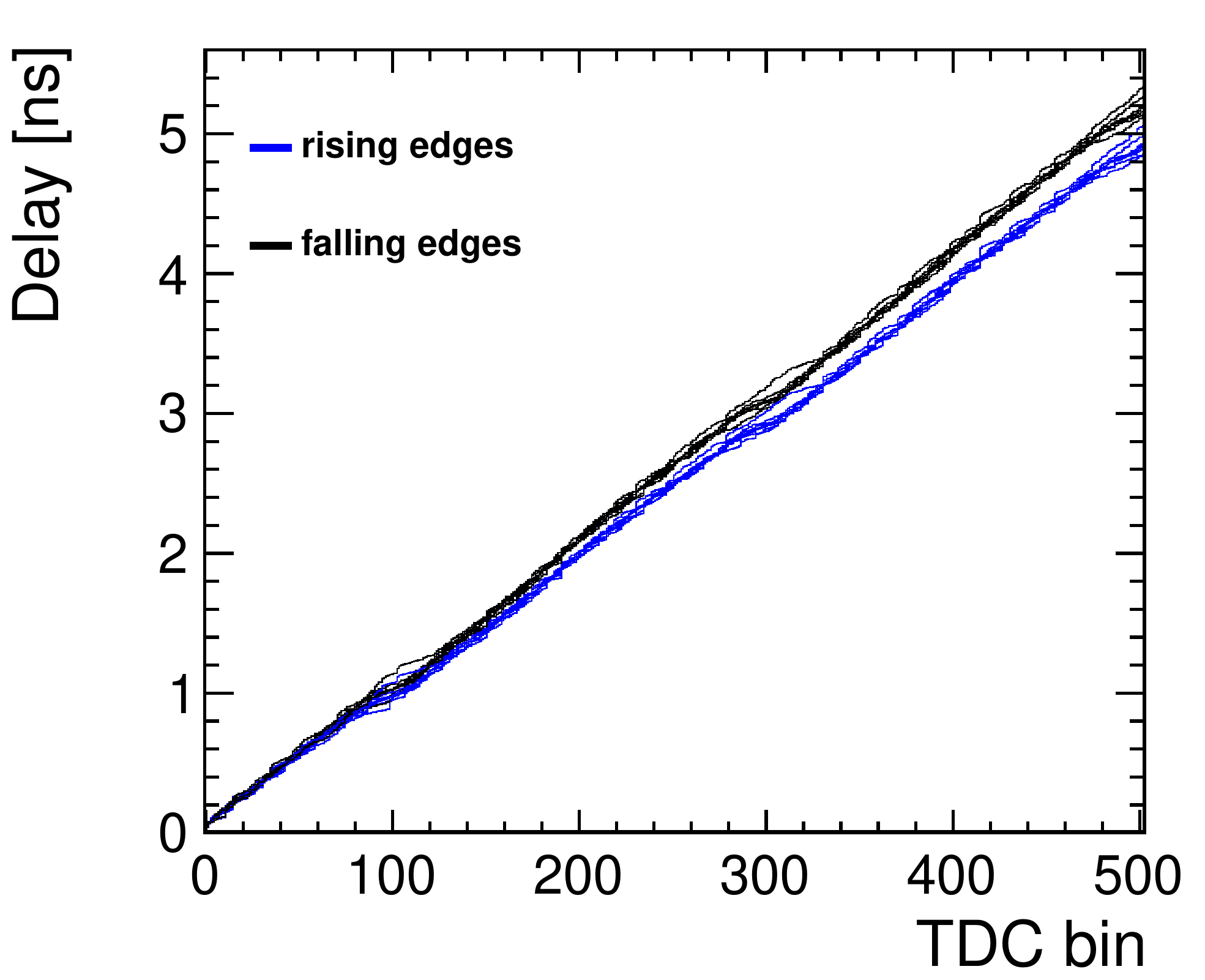}
    \caption{}
    \label{fig:lines}
  \end{subfigure}

  \caption{(a) Bin widths for rising signal edges on the first TDC channel. (b) Calibration between TDC bins and time delay for rising and falling edges.}
\end{figure}

To calibrate the TDC, a second clock independent from the TDC clock is applied to all chanels and the signal edges recorded per TDC bin are counted, which is proportional to the width of a bin.
Figure~\ref{fig:bins} shows the bin widths obtained from the calibration for rising edges on the first TDC channel. This also indicates the differential nonlinearity of the delay line, as an ideal delay line with identical bins would give a flat distribution. The average bin size is around 10\,ps with some bins ranging from 0.1\,ps to 70\,ps.
Summing the bin widths gives the calibration from TDC bins to time delay, which is shown in Figure~\ref{fig:lines}. This calibration is also used for the following measurements.
The measured lengths of the delay lines are about 5\,ns, which is in the expected range.
There is a visible difference in slope between rising and falling edges caused by different propagation speeds in the delay lines, wich necessitates a separate calibration for both. Additional non-linearities are visible at around bin 100, 300 and 500, which correspond to the delay lines crossing between different clock regions.
For rising edges, one clock cycle at 311.1111\,MHz corresponds to 331 to 334 TDC bins, which translates to a least significant bit (LSB) of approximately 9.7\,ps to 9.6\,ps. For falling edges, this corresponds to 303 to 320 TDC bins and translates to an LSB of approximately 10.6\,ps to 10.0\,ps.

\subsubsection*{Resolution}

\begin{figure}[htbp]
  \centering
  \begin{subfigure}[b]{0.49\textwidth}
    \centering
    \includegraphics[width=\textwidth]{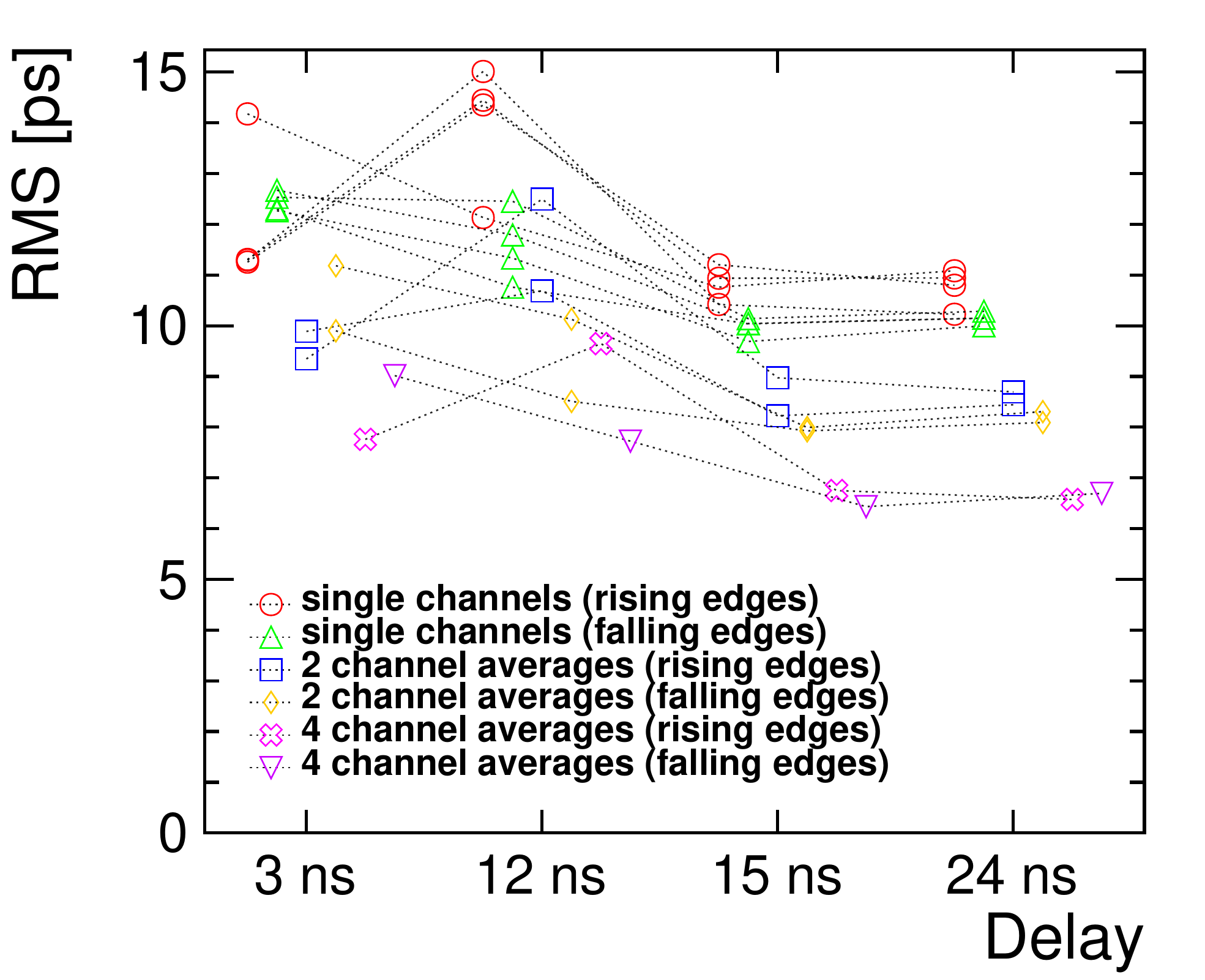}
    \caption{}
    \label{fig:resolution}
  \end{subfigure}
  \begin{subfigure}[b]{0.49\textwidth}
    \centering
    \includegraphics[width=\textwidth]{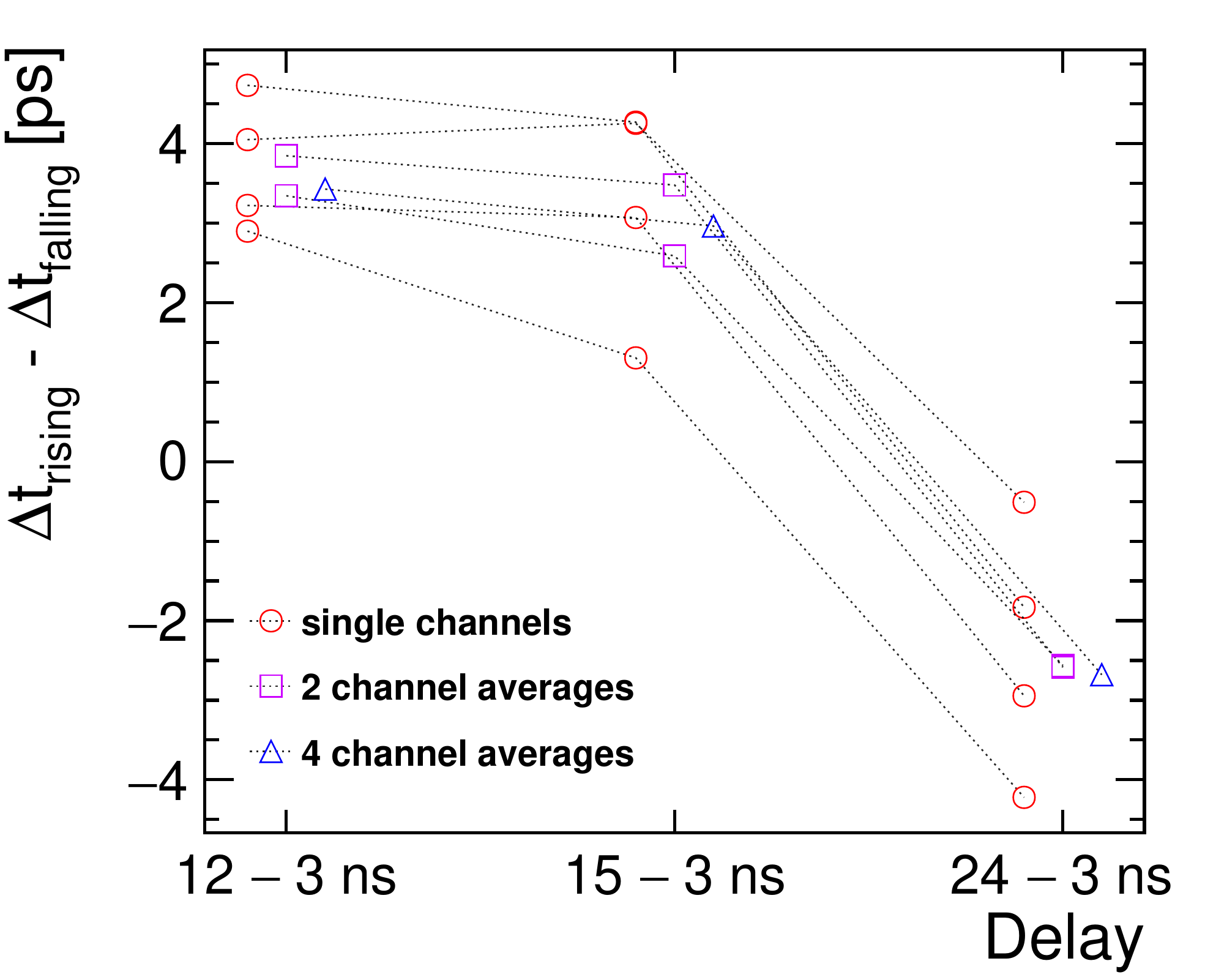}
    \caption{}
    \label{fig:deviation}
  \end{subfigure}

  \caption{(a) TDC time resolution for groups of 1, 2, and 4 channels for different signal delays. (b) Deviation of measured signal delays for rising and falling edges. }
\end{figure}

To determine the resolution of the TDC, the same calibration signal is connected directly to the first 4 channels and also split and connected to the second 4 channels with an additional cable delay of 3\,ns, 12\,ns, 15\,ns or 24\,ns, as shown in Figure~\ref{fig:caribou}. The time difference of the same signal edge between both groups is histogrammed for individual channels or averaged over groups of 2 and 4 channels and the corresponding combined time resolution $\sigma_{tot}=\sqrt{\sigma^2_{1}+\sigma^2_{2}}$ is obtained as the RMS of each distribution. About 350,000 signal edges are recorded per measurement.
The errors are estimated from the deviations between two repeated measurements and are below 0.1\,ps. 
The measured combined resolutions are summarised in Figure~\ref{fig:resolution}. Assuming similar behaviour for both groups of channels, the individual resolutions are $\sigma_{1,2}=\sigma_{tot}/\sqrt{2}$, which gives resolutions in a range of [6.9;~10.6]\,ps for groups of 1 channel, [5.6;~8.8]\,ps for groups of 2 channels, and [4.5;~6.8]\,ps for groups of 4 channels.



To test the agreement between the calibration for rising and falling edges, the measured delays for the different cable lengths are compared.
Differences in propagation speed between rising and falling edges outside of the delay lines are compensated by subtracting the measured values for the 3\,ns delay from all other measurements. This also compensates for any additional delays and differences from the routing between TDC channels.
The measured deviations between rising and falling edges for the different cable delays are summarised in Figure~\ref{fig:deviation}. In all cases, the mismatch remains below 5\,ps for all tested delays.

\subsubsection*{Linearity}

\begin{figure}[htbp]
  \centering
  \begin{subfigure}[b]{0.49\textwidth}
    \centering
    \includegraphics[width=\textwidth]{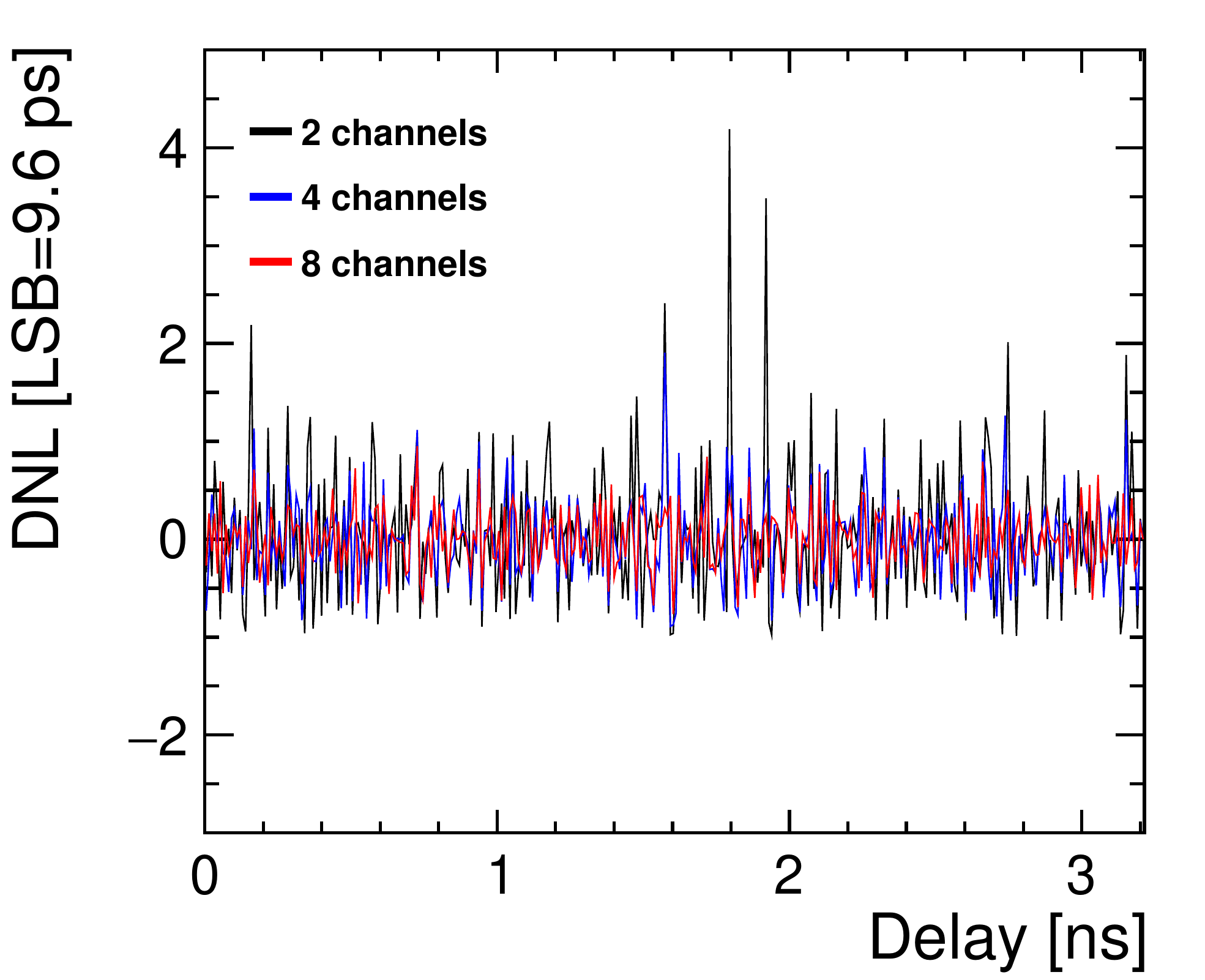}
    \caption{}
    \label{fig:dnl}
  \end{subfigure}
  \begin{subfigure}[b]{0.49\textwidth}
    \centering
    \includegraphics[width=\textwidth]{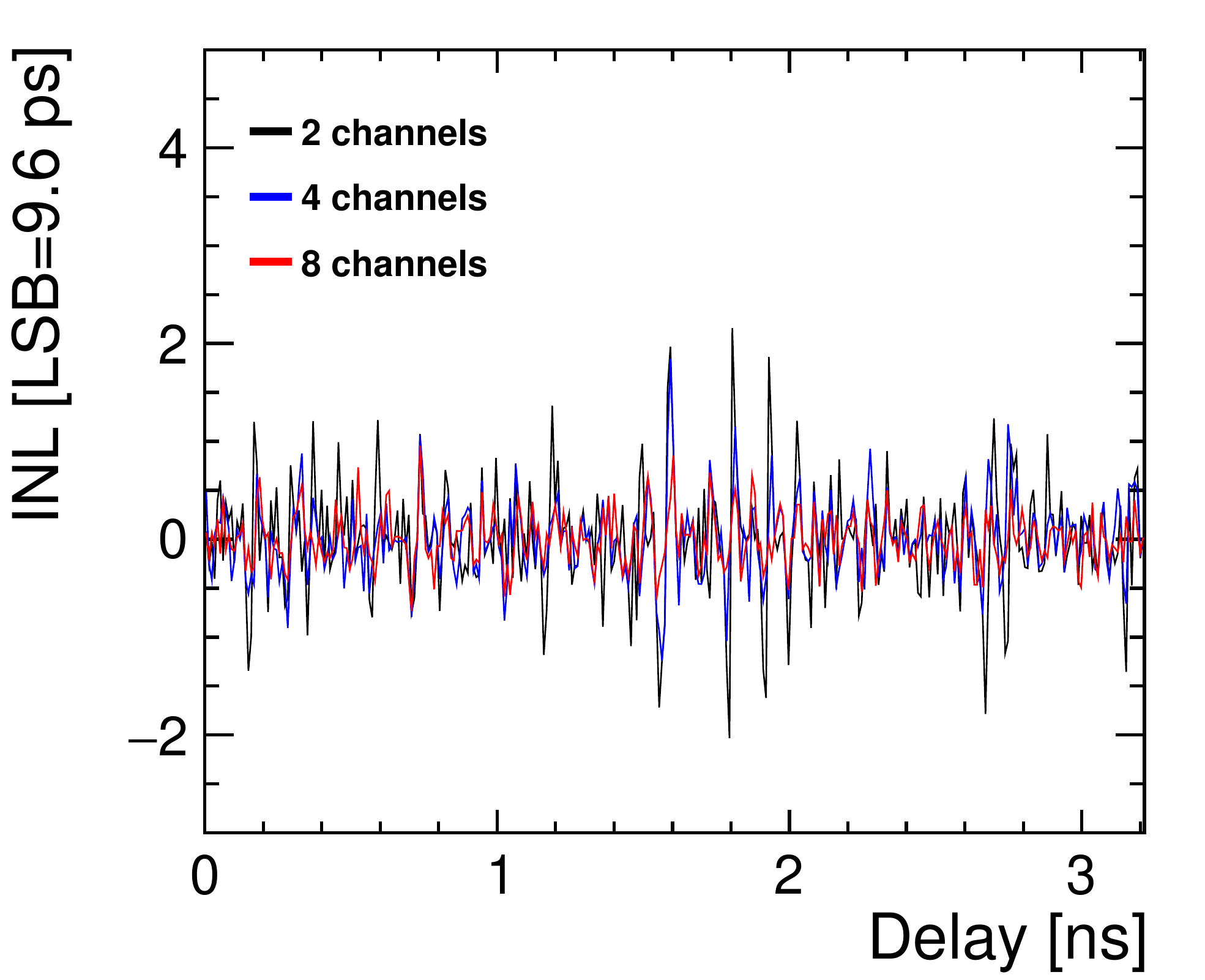}
    \caption{}
    \label{fig:inl}
  \end{subfigure}

  \caption{\label{fig:dnl_inl} DNL (a) and INL (b) after calibration for rising edges and groups of 2, 4, and 8 channels. }
\end{figure}

Figure~\ref{fig:dnl_inl} shows the differential (DNL) and integral (INL) nonlinearity for rising edges after calibration for combinations of 2, 4, and 8 channels.
With an LSB of 9.6\,ps for rising and 10.0\,ps for falling edges, the DNL for rising edges on 2, 4, and 8 channels is [-1.0;~4.2], [-0.9;~1.9], and [-0.8;~1.0]~LSB respectively. The DNL for falling edges is [-1.0;~2.5], [-0.8;~1.6], and [-0.7;~0.9]~LSB. The INL for rising edges is [-2.0;~2.2], [-1.2;~1.8], and [-0.7;~1.0]~LSB.
The INL for falling edges is [-1.4;~1.3], [-0.9;~0.9], and [-0.6;~0.5]~LSB.

\section{Summary \& Outlook}

The prototype TDC meets all requirements for the readout of the FASTPIX and DPTS and achieves a time resolution better than 10\,ps in initial laboratory tests,
which makes it a flexible alternative to oscilloscope-based readout used in previous beam test campaigns. Further tests under different operating conditions are still required to confirm the robustness of the implementation.
The TDC could be improved further by adding compression of the data stream and better matching of the delay line lengths and irregular bin sizes. 
The achieved time resolution also makes the TDC interesting for direct timing measurements by interfacing it with a dedicated timing detector in addition to the device under test.

\acknowledgments

This project has received funding from the European Union’s Horizon 2020 Research and Innovation programme under Grant Agreement No 101004761.
The work was partly performed in the framework of RD50.



\providecommand{\href}[2]{#2}\begingroup\raggedright\endgroup

\end{document}